# A Blockchain-Based Consent Mechanism for Access to Fitness Data in the Healthcare Context

**MAY ALHAJRI[1,2], (Graduate Student Member, IEEE), CARSTEN RUDOLPH[1], AND AHMAD SALEHI SHAHRAKI[1,3], (Member, IEEE)**

[1]Department of Software Systems and Cybersecurity (SSC), Monash University, Melbourne, VIC 3800, Australia
[2]Department of Computer Networks and Communications, King Faisal University, Al-Ahsa 31982, Saudi Arabia
[3]RMIT Blockchain Innovation Hub (BIH) and RMIT Centre for Cyber Security Research and Innovation (CCSRI), RMIT University, Melbourne, CBD 3000, Australia

Corresponding author: May Alhajri (may.alhajri@monash.edu)

**ABSTRACT** Wearable fitness devices are widely used to track an individual's health and physical activities to improve the quality of health services. These devices sense a considerable amount of sensitive data processed by a centralized third party. While many researchers have thoroughly evaluated privacy issues surrounding wearable fitness trackers, no study has addressed privacy issues in trackers by giving control of the data to the user. Blockchain is an emerging technology with outstanding advantages in resolving consent management privacy concerns. As there are no fully transparent, legally compliant solutions for sharing personal fitness data, this study introduces an architecture for a human-centric, legally compliant, decentralized and dynamic consent system based on blockchain and smart contracts. Algorithms and sequence diagrams of the proposed system's activities show consent-related data flow among various agents, which are used later to prove the system's trustworthiness by formalizing the security requirements. The security properties of the proposed system were evaluated using the formal security modeling framework SeMF, which demonstrates the feasibility of the solution at an abstract level based on formal language theory. As a result, we have shown that blockchain technology is suitable for mitigating the privacy issues of fitness providers by recording individuals' consent using blockchain and smart contracts.

## I. INTRODUCTION

Due to the recent growth in the use of wearable fitness devices, such as smartwatches, individuals are now exposed to vast quantities of their own sensitive health data [1]. These wearables track a variety of data, including health and physical activities such as sleep, steps, and blood pressure [2]. Although individuals, clinicians and clinical researchers benefit from adopting these technologies for accurate diagnosis, some privacy challenges have emerged [3], [4]. The European Union's (EU's) General Data Protection Regulation (GDPR) [5], which came into effect in May 2018 [5], enforces data protection regulations on data processors, giving subjects more control over their fitness data by mandating that they provide consent. The GDPR's [5] consent standards place a substantial burden on fitness providers to comply in an attempt to address the privacy concerns raised by experts [1], [6]–[13].

Consequently, several solutions that preserve privacy [14]–[16] have been offered, but they either rely on a single trusted authority or are not transparent to users, e.g., offering only one-time consent without the right to withdraw. Users expect these devices to protect their data and respect their privacy [17]. These challenges in the current wearable fitness device privacy policy urge us to further elevate the privacy level of fitness data sharing based on the user's consent. Hence, our solution brings together advances from the fields of consent management, legal frameworks, and advanced fitness data-sharing control by maintaining consent in an immutable legal archive.

This study addresses privacy issues in fitness providers' privacy policies by recording all data subjects' consent in immutable storage and giving the user more transparency about the parties with whom their data have been shared. The main goal of this research is to build a secure system model that improves data subjects' control over the processing and collocating of their data while simultaneously enabling data controllers and processors to comply with GDPR obligations. Although we can repose of the ultimate trust in a single central control system, there are associated risks of doing so. Therefore, the built-in features of blockchain make it a suitable candidate for addressing identified problems. Hence, these two contributions are based on the blockchain, which acts as a dynamic consent mechanism. The proposed approach leverages smart contracts for all automated processes, such as managing users' consent by granting or denying requests for access to their fitness data and utilizing smart contracts to detect any misbehavior by fitness providers. The following is a list of the benefits of moving fitness provider consent management to a blockchain platform:

- Transparency: All consent processes are visible to all parties, including the user.
- To improve security, blockchain can ensure that all consent-related transactions are authentic and originate from an indented agent (authenticity), have not been tampered with (integrity), consent transactions are handled in a nonrepudiable manner (proof of authenticity), and can control joint parties by choosing a permissioned network (authorization).
- Scalability utilizing blockchain can provide a scalable service to a growing number of users and nodes, including consent requests and response actions generated by different agents.
- Records' immutability can help audit records and minimize dispute cases by tracing the history of records organized in a specific sequence with timestamps. It serves as a piece of ordered evidence.
- Consent and processing/collecting fitness data validity checks are predefined in a logical manner in the form of a smart contract.

This study aims to make two contributions to the preservation of privacy among fitness tracker providers. The key contributions of this study are as follows.

1) The design and description of all aspects of the proposed framework for dynamic consent, including system design and architecture, and all potential system action sequence diagrams, show the consent-related data flow and algorithms to demonstrate how decisions are created and executed.
2) The validation of the proposed framework's security properties, including a formal abstract description of the required properties of blockchain and smart contracts, authentication, and proof of authentication, using a rigorous formal representation through the formal security modeling framework SeMF proposed by [18].

Together, these two contributions aim to address privacy issues in fitness providers' privacy policies by providing a comprehensive framework and validating the proposed system using SeMF [18].

The remainder of this paper is organized as follows. In Section II, we briefly present related work and identify the gaps related to preserving the privacy of fitness data followed by the proposed system's objectives. Section III presents the methodology used in this study. Section IV provides a refined and formal description of the system's requirements. Section V explores potential security concerns that motivate our proposed system framework and then Section VI describes the proposed system framework, including the system design and architecture, action sequence diagrams, and algorithms that demonstrate user decisions. Section VII presents our assumptions regarding the built-in security properties of blockchain. Section VIII formalizes and validates the security requirements. Finally, Section IX concludes the study with a brief discussion.

## II. RELATED WORK

The preservation of individual privacy by fitness tracker providers has been a subject of interest for many researchers. Some researchers have highlighted privacy concerns resulting from the use of fitness apps. Sunyaev *et al.*'s [13], Mulder's [10], and Hutton *et al.*'s [1] linguistic assessments have revealed a gap between users' understanding of consent and subsequent data usage by service providers. Their findings suggest the need to adopt a more transparent human-centric system that clearly communicates the purpose of requesting data. Several studies that have focused on privacy risk analysis of health and fitness app behavior suggest that fitness apps involuntarily share user data with other entities [6]–[9], [11]. Their privacy risk and behavior analysis revealed that the involuntary sharing of data is due to unclear one-time consent and the absence of withdrawal rights.

Although some fitness applications such as Fitbit [15], Apple Watch [14] and Strava [16] have moved from traditional one-time consent to more flexible user consent management, privacy concerns associated with their consent practices remain. These three Fitness applications uses a traditional data management systems that rely only on policies in their solutions to comply with data protection regulations. Hence, the current fitness applications do not technically enforce a transparency system. In addition, these applications contain no interface for informing users about third-party data access and sharing. Therefore, our proposed system uses built-in blockchain properties that technically enforce transparency in fitness applications. The aforementioned literature, which includes linguistic, behavioral, and heuristic assessments, shows serious privacy concerns associated with the use of fitness apps. Researchers have identified several problems related to preserving the privacy of fitness data, including lack of system transparency, lack of privacy policy legibility, concerns with one-time consent, and noncompliance issues with new consent management. Although these studies thoroughly analyzed the privacy issues of fitness apps, none were found to have addressed concerns in terms of dynamic consent management that uses a transparent and legally compliant system.

Since privacy is a critical component for safeguarding an individual's data, it imposes a considerable burden on service providers [19]. For instance, in the case of data sharing, data controllers and processors must adhere to appropriate standards, such as GDPR or Australian Privacy Principles (APPs) [5], [20], which require individual consent before collecting, processing, or sharing personal data. In contrast to the GDPR [5] and APP [20], which protect all forms of personal information, the Health Insurance Portability and Accountability Act (HIPAA) [21] has a limited scope. For example, fitness applications are not obliged to comply with HIPAA because they are classified as ''noncovered entities''. However, they may be required to do so once they engage with ''covered entities'' such as insurance companies and healthcare professionals [21]. While legislators enact stricter data protection laws, individuals are often overwhelmed by lengthy, complex consent forms in which they may unwittingly accept the secondary usage of their data out of bewilderment. Furthermore, individuals have no transparency in knowing if the consent terms and conditions are adhered to by service providers [21]. Thus, there needs to be a solution that provides users more control over the use of their personal data by obtaining their consent while likewise maintaining the trust of different domains and preventing unlawful secondary use (also known as ''data double-spending'') [22]. Given the rapidly changing privacy rules adopted to give individuals more control over their data, consent management can be seen as the first step for preserving privacy in any system [23].

A blockchain is a ledger with distributed ownership that has a collection of transactions included in blocks [24]. In a blockchain, the same ledger is copied and synced among peers through a consensus mechanism [25]. The blockchain holds only accepted and validated transactions encapsulated in a block [24]. Each block in the blockchain is built on top of the previous block, such that a sequence of blocks with a timestamp is created [25]. It is equipped with cryptographic techniques and is tamper-resistant, append-only, and immutable. The greater the number of confirmations the block has, the harder it becomes to change [22].

Blockchain and related technologies, such as smart contracts, have the potential to gain the trust of different domains through a predefined and automated program to satisfy common contractual conditions [24], [26]. A smart contract guarantees that it is executed immediately once a condition is met. Smart contracts employ ''if/when. . .then. . . '' rules that govern transactions beyond the control of centralized authorities. Because smart contracts operate on a blockchain, they have many valuable features, such as managing smart assets, executing as written, avoiding central entities, and reducing human interference [26], [27].

The emergence of blockchain technology has prompted the rethinking of traditional solutions that depend on trust [28]. Its unique properties, particularly immutability and decentralized authority, have attracted considerable research attention [23]. Since the initiation of Bitcoin for cryptocurrency in 2008 [29], the possibility of using blockchain in nonfinancial applications has been proposed. The use of blockchain outside of finance has been investigated in a range of contexts, such as supply chains [30]–[33] and health care and consent management [17], [34]–[38].

Many studies in various domains have empirically proven that blockchain technology can mitigate privacy preservation issues by managing consent [17], [38]–[43]. Some of these consent scheme solutions leverage smart contracts to manage individual consent. For example, the work outlined in [17] involves the use of a consortium blockchain platform in the IoT ecosystem to manage users' personal data by obtaining their consent. Another solution proposed by the Orange Consent Management Service records consent in the blockchain through a consent management server, which has been demonstrated in the Hyperledger beta version [41]. A related work proposed by Dovetail leverages a blockchain to allow patients to consent to the sharing of their electronic medical records (EMRs) with other entities through a mobile application [44]. Dovetail's blockchain stores consent along with reference to the transferred data. Similarly, an EMR management solution named Ancile, proposed by [40], utilizes smart contracts to secure access to patients' EMRs through access permissions based on HIPAA. A blockchain and smart contract-based solution introduced by [38] proposed a novel privacy-preserving dynamic consent management for EMRs secondary use, in which they used Fast Healthcare Interoperability Resources' (FHIR) consent resource in their proposed design to avoid integration failures with EMRs.

Beyond the healthcare domain, blockchain for consent management solutions has been used in other research domains. For instance, Gilda and Mehrotra [42] proposed a blockchain solution that leverages smart contracts to guarantee consent privacy and access control by establishing a nested authorization process using Hyperledger Fabric and Hyperledger Composer. Another solution introduced by [39] uses a blockchain to record user consent and enable anonymous data sharing.

To that end, utilizing a blockchain-based solution in a consent management system has great potential and offers outstanding features, such as transparency, security, immutability, and auditability [45]. Therefore, in light of the potential of blockchain and in response to privacy issues in fitness apps, this study proposes a blockchain dynamic consent mechanism that uses a smart contract to preserve privacy in accessing fitness data, especially with regard to the management of consent.

## III. RESEARCH METHODOLOGY

This section covers the research methodology and motivates the selection of the methods used throughout the research. The primary goal of this research is to improve user control over the processing of their fitness data and foster transparent data processing to achieve both privacy and compliance. This improvement is achieved by designing and describing the high-level proposed framework for dynamic consent and validating the framework's security properties, which include a formal abstract description of the required properties of blockchain and smart contracts.

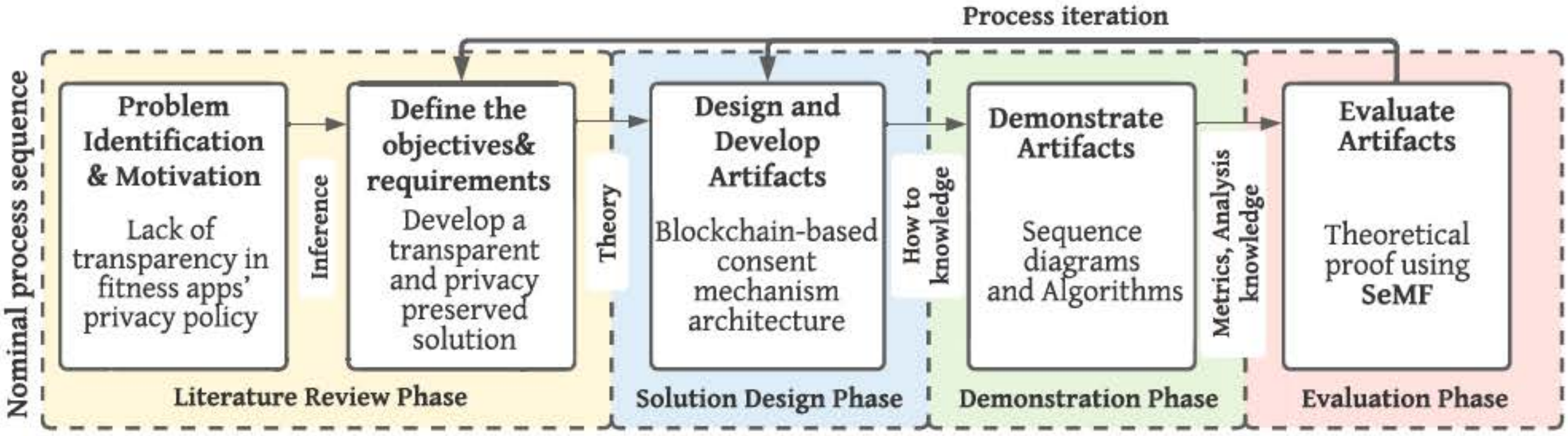

**FIGURE 1.** DSR paradigm [46], of blockchain-based consent mechanism.

We employed the design science research (DSR) paradigm proposed by Peffers *et al.* [46] to design and evaluate our framework for dynamic consent. The goal of DSR is utility, which implies that rather than researching an already existing artifact, it involves the discovery of a highly relevant problem through an iterative process of developing and assessing solution objects. This research makes use of the DSR approach to solve the issue of fitness providers' privacy because no existing solution has met all of our requirements; therefore, it involves an innovative discovery of system artifacts and evaluates them in formal abstract description using SeMF [18]. Hevner *et al.* [47] and Peffers *et al.* [46] emphasized the importance of the evaluation phase as a "crucial" component of a DSR contribution; therefore, we use SeMF, which serves as a tool for showing the trustworthiness of our proposed design model by providing an exact specification of the proposed architecture and properties and validating the architecture.

The DSR can be divided into four phases: literature review, solution design, demonstration, and evaluation, as depicted in Fig. 1. Each phase consists of a different DSR paradigm activity. Below, we describe each phase and its activities.

(1) **LITERATURE REVIEW PHASE:** In this phase, we identify the relevant problem and motivation (as explained in Section II) and define our objectives. Below are the objectives that drove our research effort:
   a) The system must closely follow GDPR valid consent criteria.
   b) Our proposed artifacts should not be allowed to overwhelm the capabilities of a typical fitness tracker's functionality.
   c) Auditing all system actions is crucial to achieving transparency and GDPR compliance requirements.
   d) In addition, there were functions and features necessary to meet transparency, including an automated self-report data access check, making data available check, and an automated switch from valid to invalid consent based upon user decisions.
   e) The proposed design for storing and exchanging user consent data between different agents must be scalable.
   f) A security solution is one of the main requirements implemented in system design to provide secure and reliable communication.

(2) **DESIGN AND DEVELOPMENT PHASE:** This phase involves the use of objectives from the literature review phase to define the required design artifacts, which involves an iterative process between the created IT artifacts and the defined objectives. The artifacts in the proposed system are as follows: (1) *system architecture* (Section VI-A), (2) *requirements specification* (Section IV) and (3) *formal model and proof* (Section IV).

(3) **DEMONSTRATION PHASE:** This phase aims to demonstrate the utility of the design artifacts in solving our identified problems and then obtain preliminary results (as explained in detail in Section VI. This phase involves the implementation of (1) *sequence diagrams*, (2) *detailed descriptions of activities*, and (3) *algorithms*.

(4) **EVALUATION PHASE:** We evaluated how well the demonstrated artifacts supported our identified problem. This phase involves comparing the results from the demonstrated artifacts with the defined objectives. The evaluation is based on *using SeMF and proof* (as explained in detail in Section VIII). This process iterates back to refine the designed artifact and improve its effectiveness.

## IV. REQUIREMENTS SPECIFICATION

As a subsequent iterative step of the DSR paradigm [46], this section provides a refined and formal description of the system's requirements from the problems and objectives identified in Section II, which is used in Section VIII for formal validation. Therefore, it is important to understand the requirements of the proposed system. The proposed solution requirements and the choices made provide the rationale for the model and mechanism properties of our proposed system, which are discussed in detail in Sections VI-A and VI-B. The notations used throughout this study are listed in Table 1.

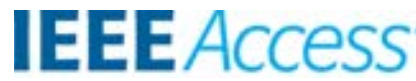

**TABLE 1. List of notations and their descriptions.**

| Notation | Explanation |
|---|---|
| $U_i$ | the individual user and the owner of fitness data. |
| $RF_i$ | unique reference number for each consent. |
| $R_i$ | an agent in the system who requests consent to collect data, known as requester or collector. |
| $FP_i$ | a set of agents in a system who request consent to process data, known as a processor or fitness provider. |
| $RA$ | an agent in a system who governs the network, known as the regulatory authority. |
| $pk_i$ | public key of *i* where *i* represents an agent. |
| $sk_i$ | private key of *i* where *i* represents an agent. |
| $D_i$ | the actual *Data* being sent or received. |
| $A_i$ | the action or transaction performed by agents that have some data $D_i$, where *i* represents the transaction's index. |
| $t_v$ | refers to the time which denotes the consent decision valid for a period of time specified by $U_i$, where *v* represents the validity of the grated consent. |
| $\lambda_P$ | an agent's local view of the system. |
| $\omega$ | the sequence of actions. |
| $\Gamma$ | the set of actions. |
| $\mathbb{P}$ | the set of all agents. |

This section discusses each of these requirements as follows: **R1:** level of transparency, **R2:** security requirements, **R3:** scalability, **R4:** auditability, **R5:** preservation of the original functionality of the fitness tracker, and **R6:** GDPR compliance.

### A. R1: TRANSPARENCY

Transparency can be defined as a way of completing an agent's activities openly without any hidden activities such that users can see their data flow and trust that the agent is fair and honest. In any system, transparency is crucial to achieve both privacy and compliance. To enhance users' trust regarding the protection of their privacy, the GDPR [5] and APPs [20] ensure transparency in data processing practices. Three fundamental pillars of transparent systems can be defined: transparency, privacy, and compliance. Users must be provided transparency regarding what happens after they consent to the processing or collection of their data. To address this requirement, the proposed system ensures that all three fundamental pillars of a transparent system are considered when fitness providers share data-processing logs with users who have consented to the process.

### B. R2: SECURITY

Security is an important part of the proposed system requirements and comprises a set of security properties that ensure the trustworthiness of our proposed system. This section presents the basic concepts of properties relevant to this study. Table 2 lists the security requirements of the proposed system. In Section VIII, we formalize the requirements in detail using SeMF.

**TABLE 2. Security properties and description.**

| Security property | Description |
|---|---|
| *Confidentiality* | Only specific agents are enabled to know the value of data *D*. To ensure that *D* is confidential, *D* cannot be disclosed to unauthorized parties or malicious agents if such disclosure might affect the privacy of $U_i$, $FP_i$ or $R_i$. *D* can include entity identities (identifiers), *D* in combinations that might reveal entity identities (quasi-identifiers), sensitive or generic *D*, and sequences of action $A_1 ... A_n$. |
| *Authentication* | This ensures that the communicating agents are who they claim to be by verifying their identity. Communication usually occurs in the form of a set of actions $A_1 ... A_n$. Thus, each time action $A_i$ occurs, it must be authenticated to the receiving agent that the action originated from that sender. For example, each time an agent receives data *D*, it must be authenticated to the receiving agent such that data *D* is indeed equal to the original data *D* sent by the sending agent. |
| *Nonrepudiation* | This denotes that the system should ensure a level of protection against any denial by any agents performing an action $A_i$ or a set of actions $\Gamma$. The system should provide proof of authenticity that all actions $A_1 ...$ $A_n$ originate from the intended agent. Nonrepudiation involves two communication pairs. First, a sent transaction cannot be denied. For example, agent $A$ sent the transaction to agent $B$; thus, agent $A$ cannot deny the sending behavior. Second, a received transaction cannot be denied. For example, agent $A$ sends the transaction to agent $B$; thus, $B$ cannot deny that it received the transaction. |
| *Integrity* | This denotes that the data *D* received by the receiving agent are equal to the data *D* provided by the sending agent. Any data *D* transferred during the occurrence of actions $A_1 ... A_n$ must not be changed during the transfer and must be assured of being tamper-free when received by another agent. |
| *Authorization* | This denotes that the system should prevent unauthorized agents from being part of the system and performing a set of actions $A_1 ... A_n$. |
| *Availability* | This denotes that the agents in the system should always have access to the system's resources, and data *D* are needed to perform actions $A_1 ...$ $A_n$. |

### C. R3: SCALABILITY

As the proposed system relies on enhancing consent management among different agents, having a scalable infrastructure is critical for achieving system requirements. The number of consensus nodes, agents, and transactions handled per second are the three key elements that determine the scalability of a blockchain [25]. Thus, in our proposed system, we utilize blockchain to provide a scalable service to a growing number of users and nodes, including consent requests and response actions generated by different agents. Although each full node maintains a copy of the entire blockchain, only the final settlement of self-report data access and available transactions are recorded in the chain, resulting in an overall higher throughput [25].

### D. R4: AUDITABILITY

Auditability is a built-in blockchain feature that keeps all the data in a blockchain immutable and resistant to tampering.

The consent recorded in the blockchain can be used to conduct audits by the regulatory authority *RA* to resolve disputes among untrusted agents. This auditability feature can also be used to prove the authenticity or nonrepudiation security requirements, which are discussed in detail in Section VII. Although auditability features might be violated due to a 51% attack in a public blockchain, the proposed system uses a permissioned network on which only authorized parties are allowed to validate blocks, with some fairness among the parties, depending on the choice of consensus algorithm (e.g., PoW, PoS, Byzantine fault tolerance, etc.).

### E. R5: PRESERVE FITNESS TRACKER ORIGINAL FUNCTIONALITY

This is important because as fitness trackers rely heavily on fast processing, the proposed solution should not hinder or burden the original functionality of the system. Although GDPR imposes the need to obtain users' consent to process their fitness data, it is not sufficient to ask for users' consent every time the data are processed and recorded in the chain. Therefore, regardless of the privacy complexity it adds to our proposed system, we ensure a sufficient level of functionality by creating a flexible contract (e.g., *ContractToProcessDataForTimePeriod()* for one month or year) and giving users the choice to revoke that contract at any time using *ConsentWithdraw()*. Therefore, the privacy requirement of the proposed system is met, and the processing operation will not be sufficiently slowed such that individuals are deterred from using the system.

### F. R6: GDPR COMPLIANCE

Consent is considered valid under the GDPR [5] if it fulfils certain criteria: it must be unambiguous, informed, freely given, specific and auditable, withdrawable, and explicit [48], [49] (as listed below).

- ***Unambiguous***. *Consent must be given by means of 'a clear affirmative action' (e.g., preselected options are invalid).*
- ***Informed***. *The subject of the data must be aware of all the information associated with processing their data before they are processed.*
- ***Freely given***. *On a voluntary basis, without coercion. The subject of the data must be aware of all the effects of the consent.*
- ***Specific***. *Consent's request must be granular with a specified purpose. Thus, the subject of the data must be fully aware of the purposes and methods used to process their data.*
- ***Auditable***. *All consent information must be stored for future auditing and can later be used as valid proofs.*
- ***Withdrawable***. *Consent requests should include a way to easily withdraw granted consent.*
- ***Explicit***. *Consent should verify that it was given by the subject of the data and should include detailed information on the data to which the consent applies. It must be verifiable to be validated.*

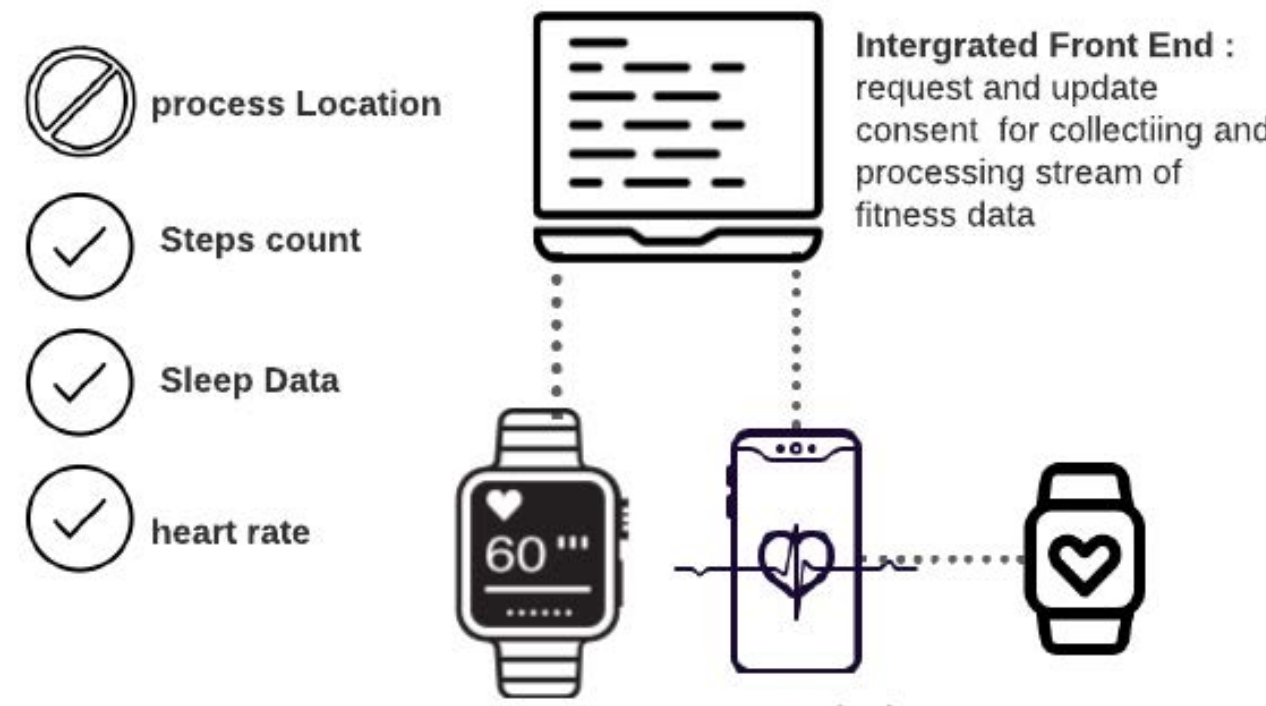

**FIGURE 2.** Device privacy settings.

## V. THREAT MODEL

A threat modeling phase to explore potential security concerns is the best practice for designing a secure system. Using such a model enables us to select countermeasures during the design process of our framework model. As the privacy policies of fitness providers present a variety of security threats, we concentrated on three types of threats: consent tampering, consent fabrication, and the unlawful processing/collection of fitness data without individual consent.

## VI. PROPOSED SYSTEM MODEL

In this section, we introduce our system architecture design model for a blockchain-based consent mechanism to access fitness data. Our proposed design focuses on privacy in terms of the exchange and records all consent processes in the blockchain. This works as a proof of consent by leveraging built-in security properties of the blockchain. The three main functionalities of the system are recording consent requests and responses, data transmission logs, and user device privacy settings. These three main functionalities ensure that the system is in compliance with the GDPR requirement, as discussed in Section IV, by tracing all consent logs and preventing any potential malicious behavior on the system when the app runs in the phone's background. One way to prevent such occurrences (as shown in Fig. 2) is to record the user's device privacy settings in the blockchain, which ensures that all data acquisition is performed in a legitimate manner based on both user consent and device settings. The device settings serve as an additional privacy-preserving layer above the user's consent level.

The proposed system acts as a legitimate legal archive in which all consent and data-sharing-related activities are recorded on the blockchain for auditing purposes. In the case of dispute/conflict or noncompliance, this process automatically forces both fitness providers and data requesters to comply with the legal framework when they acquire consent from users to process or collect their fitness data. GDPR fosters transparent data processing to achieve both privacy and compliance. In line with GDPR compliance, the proposed system provides transparency by sharing data transaction logs back to the user to give the user a transparent view of their data transmission and know more about the parties with whom their data have been shared and whether the data have been shared based upon granted consent. This approach ensures that every action performed by the entities in the proposed system remains transparent to all parties involved in the blockchain network. Although exchanging the data through blockchain provides a valuable solution, it creates a point of conflict between blockchain and one of the GDPR criteria, that is, "right to be forgotten". The immutability feature of blockchain makes it impossible for data to be erased and reverses the concept to "right to be *never* be forgotten". Although individual privacy is a fundamental right, auditing mechanisms are essential to ensure that rules and regulations are obeyed. Therefore, the proposed system was built based on the idea of decoupling consent management from data sharing. Neither the data nor the pointer of the data are stored in the chain. Blockchain only manages and stores consent and data-sharing-related activities and does not exchange data through the blockchain. The blockchain operates as a middle consent management layer between the requester and the owner of the fitness data. Blockchain stores the following data: consent requests, consent responses (grant/denial), consent withdrawal, data transmission logs, legitimate report checks of data transmission, and user device privacy settings.

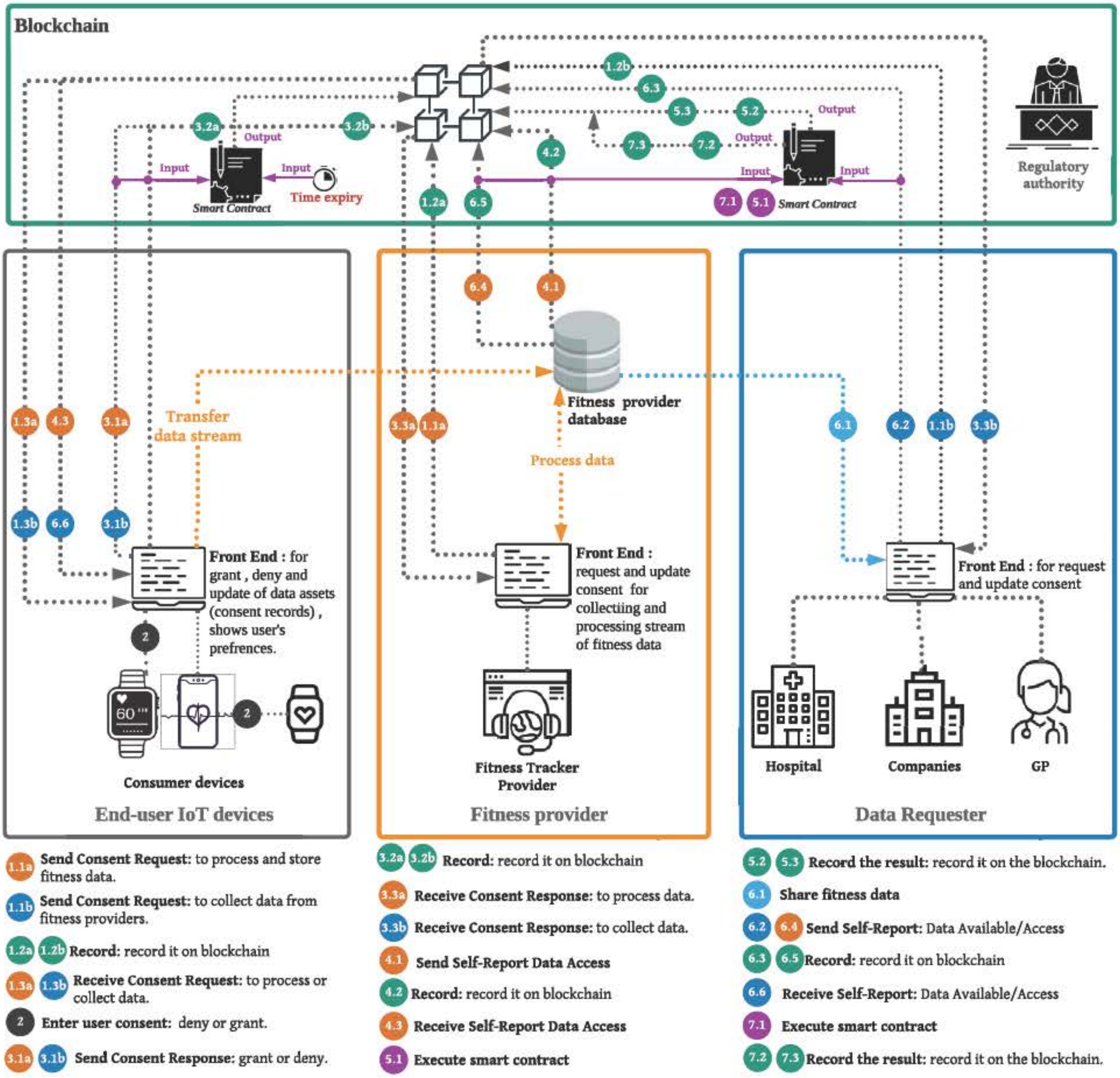

**FIGURE 3.** Proposed consent management system architecture.

### A. SYSTEM ARCHITECTURE

This section presents our design of the proposed framework for dynamic consent, including the system design and architecture, and all potential system action sequence diagrams that show the consent-related data flow and algorithms to demonstrate how decisions are created and executed. We consider the following system participants: fitness provider ($FP_i$), data requester ($R_i$), regulatory authority ($RA_i$) and user ($U_i$). As shown in Fig. 3, the proposed system is composed of four main entities that contribute differently to the system. Here, we analyze their local view, which is used later (in Section VIII) for the security proof.

- **Fitness Provider ($FP_i$):** The local view of $FP_i$ is the send and receive actions. $FP_i$ has three sending activities: consent requests to process data, self-report data access, and self-report making data available. In addition, $FP_i$ has one receiving action, which is a consent response to the process data.
- **Data Requester ($R_i$):** The local view of $R_i$ is the send and receive actions. $R_i$ has two sending activities: consent requests to collect data from $FP_i$ and self-report data access. In addition, $R_i$ has one receiving action, which is the consent response to collect data.
- **User ($U_i$):** The local view of $U_i$ is the send and receive actions. $U_i$ has three sending activities: consent responses to collect/process data and withdrawal of valid consent action. Additionally, $U_i$ has five receive actions: two consent requests from $FP_i/R_i$, three self-report data accesses or making data available to ensure transparency with the user. All the above actions are recorded in the blockchain to ensure secure and reliable communication among entities.
- **Smart Contract ($SC_i$):** The role of smart contract $SC_i$ is to automatically check the validity of the processing and collecting actions of $FP_i/R_i$. It is also used to automatically switch from valid to invalid consent based on the user's withdrawal action and expiration time.
- **Regulatory Authority ($RA$):** $RA$'s role is to ensure that the authentication information provided by other entities is lawful. When entities first register on the blockchain platform, their identity is checked by the $RA$ in which they enter their identity authentication details. The $RA$ is the first line of defense to ensure the security of the proposed system. Therefore, unauthorized nodes cannot tamper with consent, forge transactions, or attack smart contracts.

### B. PROPOSED SYSTEM SEQUENCE OF ACTIONS AND ALGORITHMS

The proposed system consists of a set of consent-related actions $\Gamma$ that are listed in Table 3 as *(Action, Entity, Parameters)*. The entities $FP_i$, $R_i$, $U_i$ and $RA$ are identified as agents acting in the system. To formalize the actions depicted in Fig. 3, we abstract the way in which fitness providers process the fitness data $D_i$ measured by user's $U_i$ wearable fitness devices and how the data are shared and processed by other agents based on $U_i$'s consent. Hence, we use the actions listed in Table 3.

**TABLE 3. Actions for proposed system.**

| Actions ($A_i$) | Explanation |
|---|---|
| **Consent Request** | |
| (SendConsentRequest,$FP_i/R_i$, (ConsentToProcess/Collect, $RF_i$)) | $FP_i/R_i$ sends a request for consent to process data. |
| (RecvConsentRequest, $U_i$,($FP_i/R_i$, ConsentToProcess/Collect, $RF_i$)) | $U_i$ receives and processes the request sent by $FP_i/R_i$. |
| **Consent Response** | |
| (SendConsentResponse, $U_i$, (ConsentDecision, $RF_i$)) | After receiving a request, $U_i$ makes a decision and sends it to $FP_i/R_i$. |
| (RecvConsentResponse, $FP_i/R_i$, ($U_i$, ConsentDecision, $RF_i$)) | $FP_i/R_i$ receives and processes the decision made by $U_i$. |
| **Self-Report** | |
| (SendSelfReportDataAccess, $FP_i/R_i$, (Processing/CollectingActivitiesUpdates, $RF_i$)) | If consent is granted by $U_i$, $FP_i/R_i$ updates $U_i$ with all activities associated with the consent $RF_i$. |
| (RecvSelfReportDataAccess, $U_i$, ($FP_i/R_i$, Processing/CollectingActivitiesUpdates, $RF_i$)) | $U_i$ receives and processes the updates sent by $FP_i/R_i$. |
| (SendReportDataAvailable, $U_i$, ($FP_i$, CollectingActivitiesUpdates, $RF_i$)) | If consent is granted by $U_i$, $FP_i$ updates $U_i$ with all collecting activities associated with the consent $RF_i$. |
| (RecvReportDataAvailable, $U_i$, ($FP_i$, CollectingActivitiesUpdates, $RF_i$)) | $U_i$ receives and processes the updates sent by $FP_i/R_i$. |
| **Withdraw Consent** | |
| (SendWithdrawConsent($FP_i/R_i$, $U_i$, (WithdrawConsent, $RF_i$)) | If $U_i$ sends consent withdraw to $RF_i$. |
| (RecvWithdrawConsent, $FP_i/R_i$, ($U_i$, WithdrawConsent, $RF_i$)) | $FP_i/R_i$ receives and processes the withdraw request sent by $U_i$. |

Each agent in the system has its own local view denoted by $\lambda_P$, in which an agent can see only its own sent actions or received actions $A_1 \ldots A_n$. We can summarize each agent $P$'s $\lambda_P$ by identifying the send and receive actions performed by each agent $P$. Hence, the design goals of the proposed system can be informally stated as follows.

**DG**1 *It must be authentic for an agent that the data shown on the agent device are the data transferred in the blockchain.*

**DG**2 *It is authentic for agent B that the data received are the data that were sent by agent A.*

The blockchain feature allows all actions $A_1 \ldots A_n$ performed by agent $P$ to be recorded on the chain and viewable to all agents $\mathbb{P}$. This process can enable agent $P$ to learn more about other actions and expand their $\lambda_P$ of system actions beyond sending and receiving actions. All the system actions $A_1 \ldots A_n$ are listed in Table 3. Based on the aforementioned assumptions and our knowledge of blockchain mechanisms such as immutability and smart contracts, we assume that they can be achieved through smart contracts.

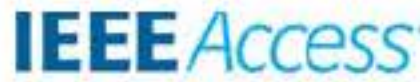

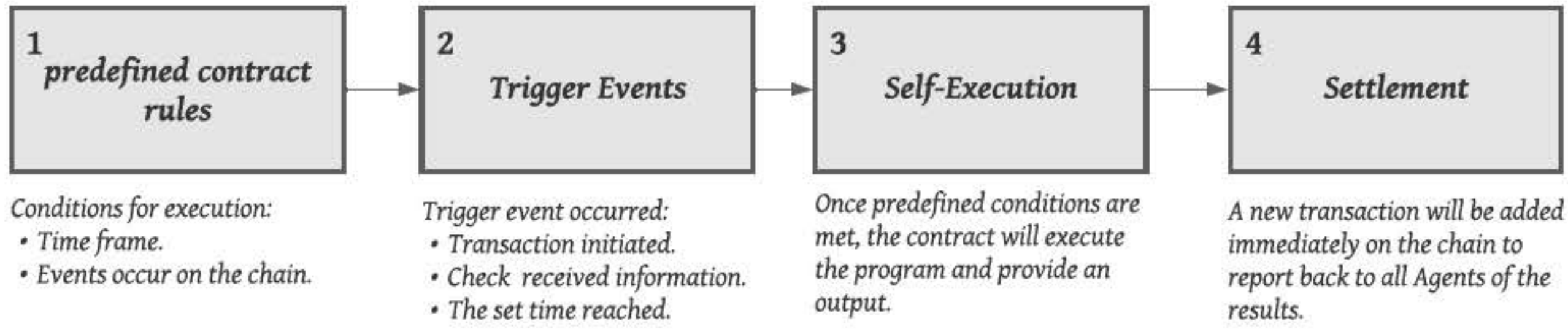

FIGURE 4. Smart contract execution steps.

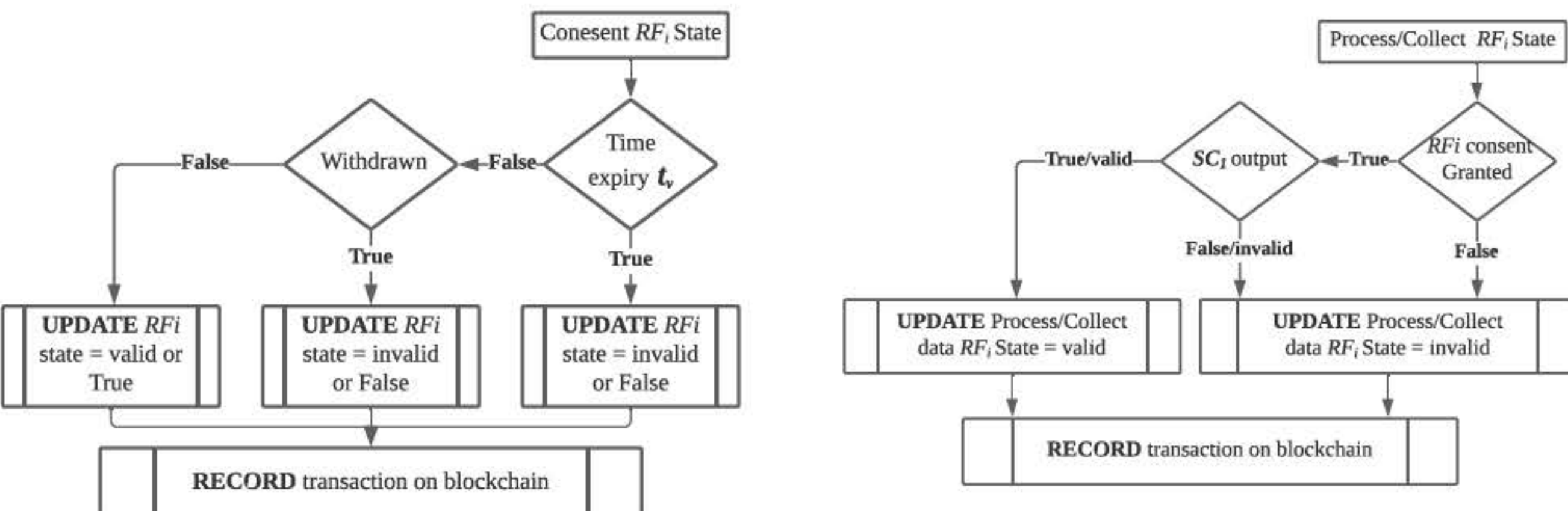

FIGURE 5. Smart contract $SC_1$ logic model.

FIGURE 6. Smart contract $SC_2$ logic model.

Fig. 4 shows an illustrative diagram of the smart contract $SC_i$ execution steps.

In the proposed system, we have two automated processes in which we can employ our assumptions about smart contracts. These automated processes are as follows.

(A) ***Smart contract*** $SC_1$- Granted consent decision is valid for a period of time ***OR*** invalid if it has been withdrawn:
A smart contract enables us to bind time with granted consent by $U_i$. Once the time has expired, the granted consent becomes invalid (as shown in Fig. 5). Therefore, the processing and collection of data by $FP_i$ or $R_i$ after that set time would be unlawful. The smart contract $SC_1$'s input parameters are $SC_1$'s time expiry $t_v$ and the initiated withdrawal transactions from $U_i$. The output of $SC_1$ adds a new transaction to the blockchain to report the consent with that reference number $RF_i$ becoming invalid.
Below, we describe design goals based on the properties of $SC_1$ that provide authenticity to agents, which can be informally stated as follows (as formalized in Section VIII):

**DG3** *It must be authentic for all agents that the granted consent is valid upon both the consent timeframe and $U_i$'s withdrawal decision.*

(B) ***Smart contract*** $SC_2$- Check the validity of storing, processing, and collecting data based upon $U_i$'s granted/denied consent:
A smart contract enables us to detect any misbehavior from $FP_i$ or $R_i$ in processing, storing, and collocating $U_i$'s fitness data (as shown in Fig. 6). Misbehavior is detected through a conditional check. Once a triggering action is recorded in the blockchain, its validation and authenticity are checked based upon $U_i$'s consent decision (granted/denied) previously made for that consent with $RF_i$. $SC_2$ depends on three input parameters: the initiated transactions from $FP_i$ or $R_i$, the output of $SC_1$, and the recorded $U_i$'s denial consent. The identifier of consent is the $RF_i$ reference number.
Below, we express the design goals informally based on the properties of $SC_2$ (as formalized in Section VIII):

**DG4** *The $FP_i$ processed data must be authentic for $U_i$ based upon $U_i$'s granted consent.*

**DG5** *The $R_i$ collocated data must be authentic for $U_i$ based upon $U_i$'s granted consent.*

After agent $P$ joins the system, its authenticity is verified by the system's regulatory authority $RA$. The agent can start using the blockchain to record consent actions by sending a consent request to user $U_i$, the owner of the fitness data, and waiting for granting or denial of the consent request.

Below, we define all possible steps of the consent request/response process (as shown in Fig.7). The sequence diagram in Fig. 8 shows how participating agents are connected and how consent is created and recorded in the chain by showing the transaction flow among them.

- Steps 1.1a and 1.1b: $FP_i/R_i$ creates a consent request that includes three main components: consent content, reference number $RF_i$ and time required to process or collect data $t_v$. We use this information as an input for Step 2, where the user can make a decision based on these received details. This transaction is then sent to the target user through a blockchain network.
- Step 1.2: The transaction sent by $FP_i/R_i$ is recorded on the blockchain. First, a transaction or set of transactions $A_1 \ldots A_n$ are validated and propagated through the network. Second, the valid set of transactions $A_1 \ldots A_n$, which has correct signatures and structure, is enclosed in a new block. Finally, the new block is the next block to the end of the longest chain (if there are competing chains), which has already been validated and decided by nodes on the network by reaching a consensus using consensus algorithms. The consensus algorithm ensures secure and reliable communication among agents $P$.
- Step 1.3: $U_i$'s blockchain account then receives transaction $A_i$, verifies it, and displays it to user $U_i$.
- Step 2: User $U_i$ decides based on the received data $D_i$ (as shown in Algorithm 1) whether to deny or grant the consent request. The decision is based on the input from requester $R_i$ (as noted in Steps 1.1a and 1.1b). The inputs are the consent content, reference number $RF_i$, and expiration time $t_v$ of consent. The output of this step is used as an input in Steps 5.1 and 7.1.
- Steps 3.1 and 3.2: The user's consent response is returned to $FP_i/R_i$. The response is recorded in the blockchain following steps similar to those explained in Step 1.2.
- Steps 3.3a and 3.3b: $FP_i/R_i$'s blockchain account receives transaction $A_i$, verifies it, and displays it to $FP_i/R_i$.

If consent is granted in Step 2, the user has the option to withdraw consent at any time. As noted in Step 2 and Algorithm 1, the user is provided with a time expiry $t_v$ for their granted decision $RF_i$. Below, we explain all possible steps to withdraw from the granted consent $RF_i$ in line with the GDPR requirements (from Section VI). All possible withdrawal transactions steps are as follows:

- Step 1.1w: User $U_i$ decides to withdraw the consent granted in Step 2 with reference number $RF_i$.
- Steps 1.2w - 1.3w: The user's withdrawal consent is sent back to $FP_i/R_i$. The sent withdrawal transaction is recorded in the chain following steps similar to those explained in Step 1.2.

**Algorithm 1** Request Consent to Process/Collect Fitness Data

**Require:**
- Consent details:
  - Purpose of process or collection.
  - Requester information, i.e., identifier $R_i/FP_i$.
  - Requested fitness data type.
- Expiration time $t_v$.
- $RF_i$

**if** $(ConsentToProcess/Collect() = Granted)$ **then**
  $FP_i/R_i$ can start the process or collect data based on the agreed time $t_v$ after receiving a granted decision through RecvConsentResponse($b$) from $U_i$.
**return** True;
**else if** $(ConsentToProcess/Collect() = Denied)$ **then**
  The processing or collection of data after receiving a denial decision through RecvConsentResponse($b$) from $U_i$ is unlawful.
**return** False;
**end if**

- Steps 1.4wa and 1.4wb: After withdrawal, the transaction is recorded on the chain. $FP_i/R_i$'s blockchain account then withdraws transaction $A_i$, verifies it, and displays it to $FP_i/R_i$.
- Steps 2.1w and 2.2w: In these steps, the predefined smart contract $SC_1$ (defined in Algorithm 2) is executed. $SC_1$ aims to change the granted valid consent to invalid consent for processing or collecting data. The $SC_1$ program is executed if one of the following actions occurs: expiration of the granted consent time $t_v$ or withdrawal of transactions initiated by $U_i$ from Steps 1.2w and 1.1w.

**Algorithm 2** Smart Contract ($SC_1$) for Switching Consent's Validity Status

**Require:**
- The $U_i$ initiated withdrawal transactions.
- Expiration time $t_v$.
- $RF_i$

**if** $((t_v = True) \vee (withdrawal = True))$ **then**
  Inform the agents that the consent with $RF_i$ has become **invalid** by reporting the result to the chain via $SC_1$.
**return** False;
**else if** $((t_v = False) \wedge (withdrawal = False))$ **then**
  The consent with $RF_i$ remains **valid**.
**return** True;
**end if**

- Step 2.3w: $SC_1$ from Steps 2.1w and 2.2w ensures that the proposed system design goal DG 3 is held in the system. The results of $SC_1$ are reported as a new transaction (as formalized in Section VIII). $SC_1$'s transaction is recorded in the chain, following steps similar to those

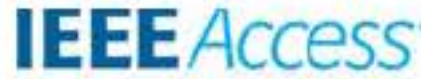

**Consent Request and Response Actions**

**Step1 Consent Request**
- **1.1a:** Send action from *FPi*
- **1.1b:** Send action from *Ri*
- **1.2:** Record action on Blockchain
- **1.3:** Receive action by *Ui*

**Step 2 $U_i$ enters Consent Decision**

**Step 3 Consent Response**
- **3.1:** Send action from *Ui*
- **3.2:** Record action on Blockchain
- **3.3a:** Receive action by *FPi*
- **3.3b:** Receive action by *Ri*

**Step 4 Self-Report Data Access**
- **4.1:** Send action from *FPi*
- **4.2:** Record action on Blockchain
- **4.3:** Receive action

**Step 5 Check the validity of processing**
- **5.1:** Excute Smart Contract $SC_2$
- **5.2:** Send results to Blockchain
- **5.3:** Record action on Blockchain

**Step 6 Report Making Data Available**
- **6.1:** *FPi* share data with $R_i$
- **6.2:** Send action from *FPi*
- **6.3:** Record action on Blockchain
- **6.4:** Send action from Ri
- **6.5:** Record action on Blockchain
- **6.6:** Receive action by $U_i$

**Step 7 Check the validity of Collecting**
- **7.1:** Excute Smart Contract $SC_2$
- **7.2:** Send results to Blockchain
- **7.3:** Record action on Blockchain

**Withdraw Granted Consent Actions**

**Step1 Withdraw Consent**
- **1.1w:** Withdraw GrantedConsent
- **1.2w:** Send action from *Ui*
- **1.3w:** Record action on Blockchain
- **1.4wa:** Receive action by *FPi*
- **1.4wb:** Receive action by *Ri*

**Step2 Update Consent Validity**
- **2.1w:** Excute Smart Contract $SC_1$
- **2.2w:** Check time expiry $t_v$ and the initiation of withdrawal action.
- **2.3w:** Send results to Blockchain
- **2.4w:** Record action on Blockchain

**FIGURE 7.** Summary of system steps labels.

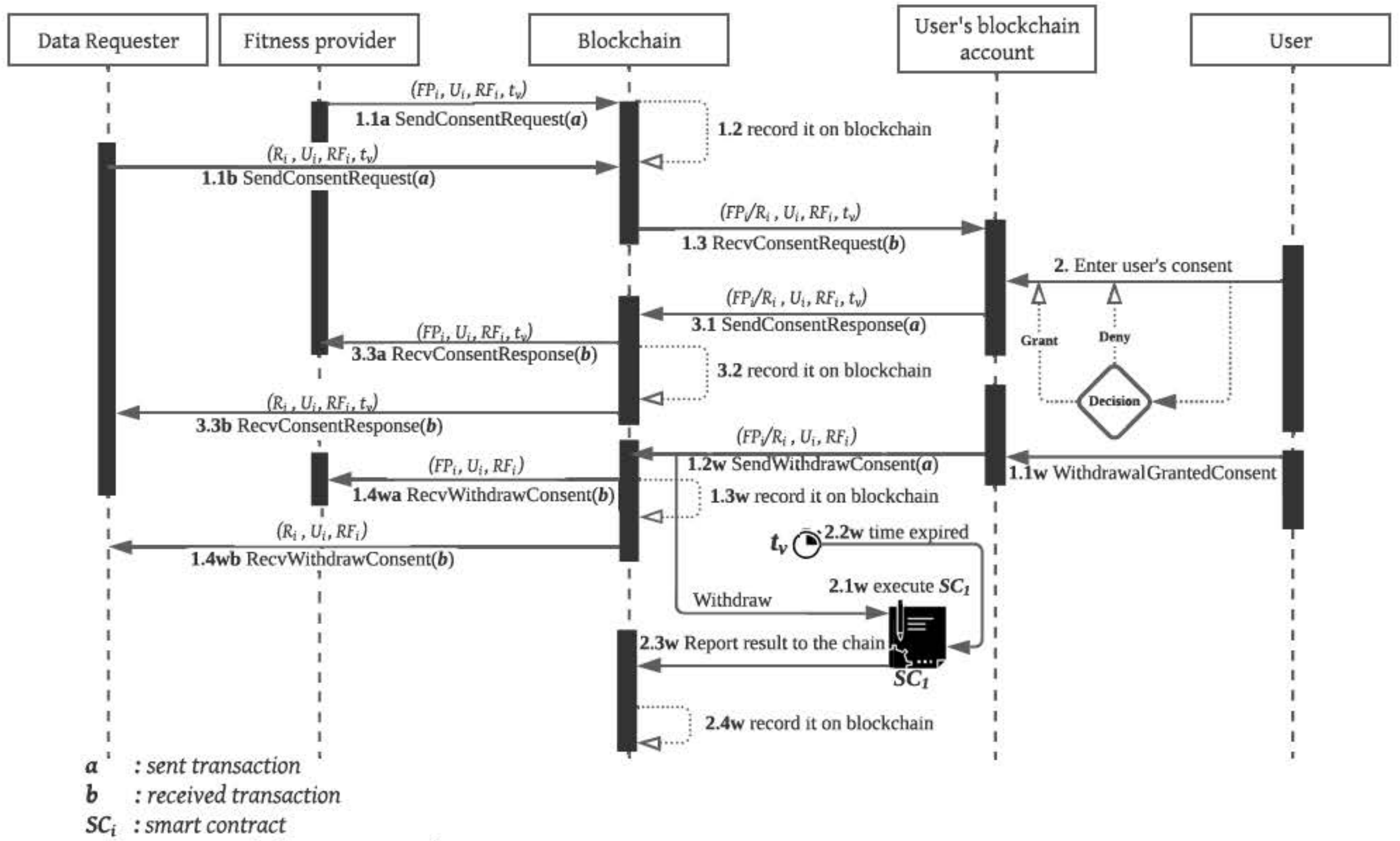

**FIGURE 8.** A sequence diagram of consent request, response and withdrawal send and receive transactions, including a smart contract $SC_1$ for switching consent validity status.

explained in Step 1.2. This step informs agents that consent to $RF_i$ has become invalid.

Fig. 9 illustrates how the proposed system's steps to check the validity of storing, processing, and collecting data are based upon $U_i$'s granted/denied consent. The following steps use the smart contract to ensure that properties DG 4 and DG 5 hold in the system. These steps aim to show that if *ConsentToProcess/Collect()* is granted, then $FP_i/R_i$ can lawfully start processing/collecting data within the agreed time $t_v$. This check uses a smart contract with predefined rules, that is, a self-executing and self-enforcing contract. Below, we define how all steps of checking the validity of storing, processing, and collecting data are based upon $U_i$'s granted/denied consent.

The sequence diagram in Fig. 9 shows the transaction flow among the participating agents (as subsequent steps of Steps 3.3a and 3.3b in Fig. 8). The steps are as follows:

- Step 4.1: Following Step 3.3a, in which $FP_i$ receives the user's granted consent, $FP_i$ starts processing and collecting data from $U_i$'s devices based upon the granted consent time $t_v$. In this stage, $FP_i$ reports back to $U_i$ that it accessed their data and processed it, with sufficient details about where and what data have been processed and collected.
- Step 4.2: $FP_i$'s self-reported data access transaction is sent back to $U_i$ and recorded in the chain, following steps similar to those explained in Step 1.2.
- Step 4.3: $U_i$'s blockchain account receives transaction $A_i$, verifies it, and displays it to user $U_i$. This received and recorded transaction is later used as an input for smart contract $SC_2$ in Step 5.1.
- Step 5.1: In this step, the predefined smart contract $SC_2$ (defined in Algorithm 3) is executed. $SC_2$ aims to check the validity of storing, processing, and collecting data by $FP_i$ based upon $U_i$'s granted/denied consent. The $SC_2$ program is executed once a triggering action from Step 4.2b is recorded in the blockchain. The validity and authenticity of action are checked based on three input parameters: the initiated transactions from $FP_i$ or $R_i$, the output of $SC_1$, and the recorded $U_i$'s granted/denied consent. The identifier of consent is the $RF_i$ reference number.
- Steps 5.2 and 5.3: The smart contract $SC_2$ from Step 5.1 ensures that the proposed system design goals DG 4 and DG 5 hold in the system and reports results of $SC_2$ as a new transaction (as formalized in Section VIII). $SC_2$'s transaction is recorded in the chain, following steps similar to those explained in Step 1.2. This step informs the agents that the storing, processing, and collecting of data by $FP_i$ are valid or invalid. This smart contract $SC_2$ acts as a detector of any misbehavior from $FP_i$ through a conditional check.
- Steps 6.1, 6.2, and 6.4: Following Step 3.3a, in which $R_i$ and $FP_i$ receive the user's granted consent, $FP_i$ shares $U_i$'s data with $R_i$. Concurrently, $R_i$ starts collecting data from the $FP_i$ database based on the granted consent and the agreed time $t_v$. In this stage, $FP_i$ and $R_i$ report back to $U_i$ that they have made data available to $R_i$, accessed their data and processed it, with sufficient details about where and what data have been processed and collected.
- Steps 6.3 and 6.5: The self-report data access by $R_i$ and report-making data made available by $FP_i$ are sent back to $U_i$ with the consent reference number $RF_i$ and recorded on the chain, following steps similar to those explained in Step 1.2.
- Steps 6.6a and 6.6b: $U_i$'s blockchain account receives transaction $A_i$, verifies it, and displays it to user $U_i$. This received and recorded transaction is later used as an input for smart contract $SC_2$ in Step 7.1.
- Step 7.1: Following steps similar to those explained in Step 5.1, predefined $SC_2$ (as explained in Section VI-B) is executed. However, $SC_2$ aims to check the validity of the data-collection process of $R_i$.
- Steps 7.2 and 7.3: Smart contract $SC_2$ from Step 7.1 ensures that DG 3 holds in the system by reporting the results of $SC_2$ in a new transaction and recording it in the chain, following similar steps to those explained in Step 1.2. This step determines whether the process of collecting data by $R_i$ is valid. This smart contract $SC_2$ acts as a misbehavior detector for both $FP_i$ and $R_i$ through a conditional check.

## VII. TRUSTWORTHINESS OF PROPOSED BLOCKCHAIN SYSTEM

As a subsequent step in the DSR paradigm, this section discusses the assumptions that model the blockchain properties, which are later validated in Section VIII. The proposed design includes four main agents $\mathbb{P}$: $RA$, $FP_i$, $R_i$, and $U_i$. Each contributes differently to the blockchain and has its own $\lambda_P$ [18]. Therefore, the trustworthiness of the system is vital to ensuring that the given consent request or response is authentic, originates from a trusted entity, and is safely stored with a backup of all data in case of failure. Blockchain can guarantee the trustworthiness of our proposed consent management system by providing an immutable and transparent way to record transactions between nontrusting entities. Entities in a blockchain have a local view ($\lambda_P$) of the system and trust only

**Algorithm 3** Smart Contract ($SC_2$) Checks the Validity of Storing, Processing, and Collecting Data Based Upon $U_i$'s Granted/Denied Consent

**Require:**
- The $FP_i$/$R_i$ initiated transactions.
- The recorded $U_i$'s granted/denied consent.
- The output of $SC_1$ (Algorithm 2).
- $RF_i$

**if** $((SendSelfReportDataAccess(a) = True) \wedge ((ConsentToProcess/Collect() = Granted) \wedge (SC_1Output = True))$ **then**

Inform the agents that the storing, processing, and collecting of data by $FP_i$/$R_i$ with $RF_i$ is **valid**.

**return** True;

**else if** $((SendSelfReportDataAccess(a) = False) \wedge ((ConsentToProcess/Collect() = Denied) \wedge (SC_1Output = False))$ **then**

Report a warning of misbehavior by $FP_i$/$R_i$ as **invalid**.

**return** False;

**end if**

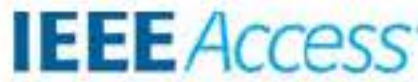

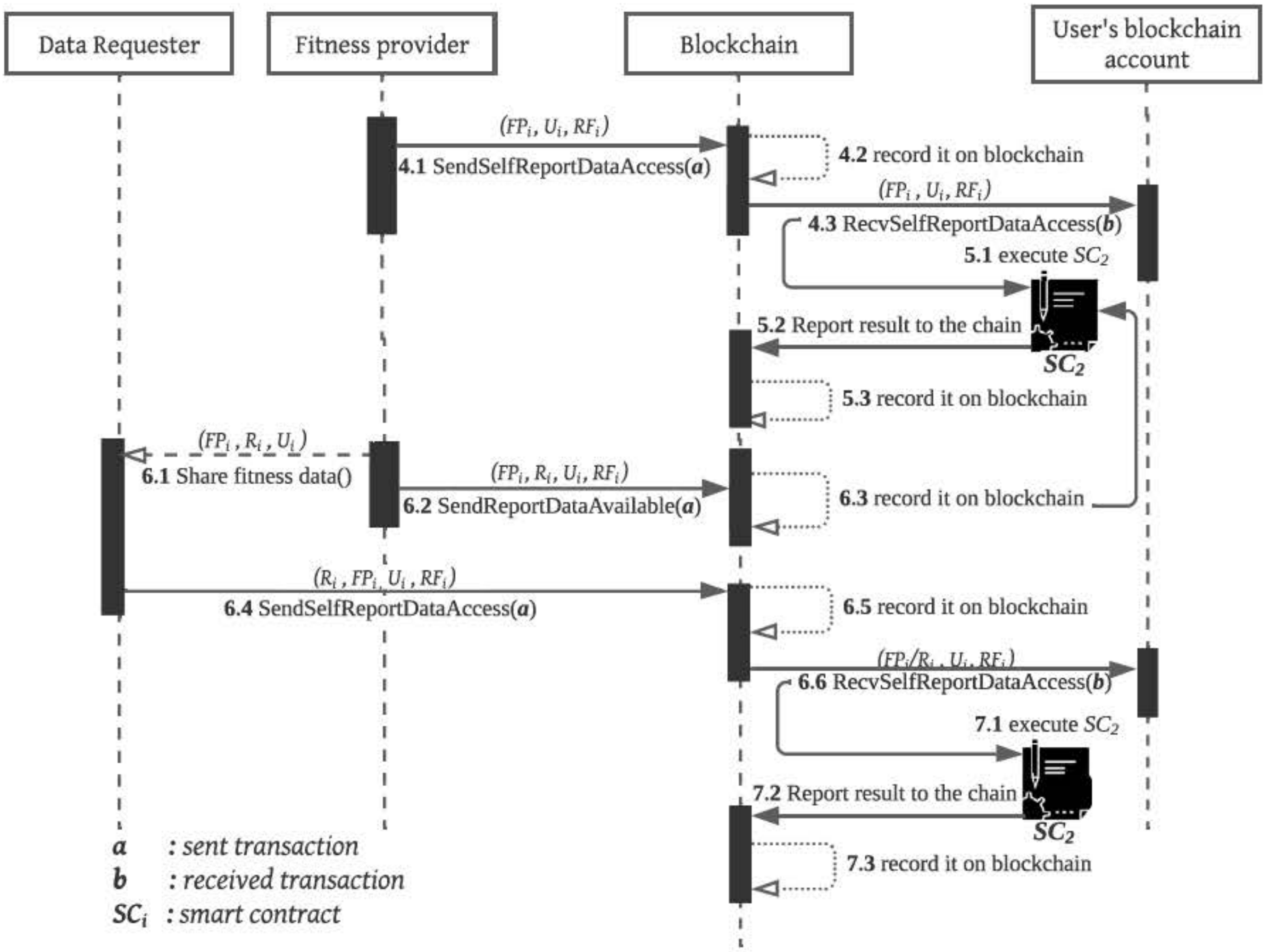

**FIGURE 9.** A sequence diagram of consent self-report transactions by $FP_i/R_i$, including a smart contract $SC_2$, to check the validity of storing, processing and collecting data based upon $U_i$'s granted/denied consent.

the results shown to them. Their trust in the shown results is based on a sequence of actions known as $\omega$ (system behavior, i.e., consent sent/received). As Fuchs *et al.* [18] noted, trust in a technological system must always be perceived as trust in a system property. The proposed system relies on assumptions about the blockchain's built-in mechanism (entity knowledge about the system). In Section VIII, we formally validate the properties of the proposed system, including its behavior (discussed in Section VI), based on our assumptions of blockchain.

This section discusses what the blockchain can do to ensure the trustworthiness of the proposed system's desired behavior. Section VIII thereafter presents clear, formal semantics that ensure the traceability of the trust and security requirements through a sequence of actions. Establishing trust between parties is critical given that trust is strongly related to security and privacy. Although we can repose trust in a single authority, there are risks associated with doing so, such as malicious acts, lack of transparency, and forged consent. Moreover, users must blindly trust a single authority to handle their data rather than trust the technology behind it. Implementing a system that uses distributed technology rather than a centralized authority can resolve trust issues. Blockchain technology introduces many built-in features that make it a suitable candidate for the proposed design. These features include authenticity, integrity, tolerance to node failure (availability), nonrepudiation, ledger immutability, and auditability.

Blockchain technology is vulnerable to potential attacks on data privacy, making it less suitable as a data-sharing platform. All transactions in the blockchain are recorded in blocks in plaintext, making it possible for sensitive information in transactions to be exposed to all agents, including opponents. Therefore, in our proposed solution, we carefully address these security and privacy issues using a blockchain as a consent management platform without sharing any data. We likewise impose encryption on some consent content, if required. We further introduce an additional security measure to increase the trustworthiness of the system by adding an access control layer to the blockchain's participants and ensuring confidentiality by using an encryption algorithm to avoid disclosing data to unauthorized agents.

The built-in cryptographic properties of blockchain provide security guarantees to any system. Each blockchain's built-in feature provides unique security properties. In Section VIII, we prove that the security requirements of the proposed system are met based on our assumptions of the blockchain's built-in security properties. Below are our assumptions, stated as formal descriptions of the model's blockchain properties.

*Assumption 1: If an action $A_i$ is received from the blockchain, it must have been previously added via a "Send" action or generated by a "Smart Contract" (smart contract/authenticity).*

*Assumption 2: If an action $A_i$ is added to the chain from a "Send" action, the agent's "Send" action must have been authentic (authenticity/digital signature).*

*Assumption 3: If action $A_i$ is added to the chain and generated by a "Smart Contract", the agent's "Send" action must have been authentic (smart contract/authenticity).*

*Assumption 4: If the data are shown on the agent's blockchain account, they must have been authentic, i.e., previously transferred in the blockchain.*

*Assumption 5: Whenever agents "Receive" data from the chain, the "Send" action adding the data to the chain must have been authentic, i.e., no data has changed (authenticity/integrity).*

*Assumption 6: A smart contract is executed, triggered, and automatically enforces the agreement terms when certain predefined conditions are met (authenticity/smart contract).*

Having evidence on record means that actions shown must have occurred sometime in the past in a specific sequence and originated by the intended agent, thereby ensuring authenticity. Below, this property of the blockchain is described in Assumptions 7 and 8.

*Assumption 7: If action $A_i$ is added to the chain, $A_i$ is enclosed in a new block and added to the end of the longest chain, after which the blocks are organized in a specific sequence of order (timestamp/order evidence).*

*Assumption 8: A copy of the whole blockchain and its valid transactions are maintained and propagated through the network nodes (immutability/evidence record).*

## VIII. FORMALIZATION OF SECURITY REQUIREMENTS

We use SeMF, developed by [18], for formal system modeling and validation of the security properties of the proposed system. This framework is based on the foundation of formal language theory, which employs fine-grained notions to model the requirements and properties of the proposed system. A detailed description of SeMF, which is beyond the scope of this study, can be found in [18]. In this section, we use SeMF's formal definitions of the security requirements to formalize some of the proposed system model requirements. These security requirements are ***(A)*** *authentication* and ***(B)*** *proof of authenticity* (with *integrity* and *authorization* properties implied in the discussion).

### A. AUTHENTICATION

#### 1) DEFINITION OF THE PROPERTY

Authentication denotes that the communicating agents are who they claim to be by verifying their identities. We use Definitions 1 and 2 from SeMF [18] to formalize the authenticity property introduced in Section IV.

#### 2) FORMALIZE AUTHENTICITY OF ACTIONS

The blockchain type in the proposed system is a permissioned blockchain in which participants are added by *RA*. The process of adding participants is outside the proposed system model and involves *RA* verifying their actual identity and then issuing key pairs for blockchain accounts. In the proposed system model, we assume that all participants were authenticated and verified by the *RA* before they became part of the system. Whenever an agent wants to add something to the blockchain, the first transaction is signed with $sk_P$. Later, the transaction is validated by the blockchain using its $pk_P$. Therefore, the system does not need a global public key infrastructure (*PKI*), where everybody trusts the key; only one agent, which is not a single entity but a distributed system (blockchain), checks the authenticity. Assumptions 7 and 8 enable all agents to know whether any action in the system is authentic and originates from the intended agent.

#### 3) END TO END AUTHENTICATION TRUST

Agent $P$, as a person, is not able to validate a digital signature, which usually requires an end device or accounts to perform the validation. Therefore, the agent accounts, not the agents $P$ or their end devices, validate the authenticity. Based on Assumption 4, the entity in the proposed system that performs the authentication validation check is a blockchain network. Therefore, the blockchain accepts only a signed valid transaction with the correct $sk_P$ and checks the origin of the transaction. For the blockchain to verify that the added action $A_i$ by agents $P$ on the blockchain is authentic and in fact originated by that agent, every agent $P$ in the system has a blockchain account or set of accounts that are composed of $pk_P$ and $sk_P$ pairs [25], as depicted in Fig. 10. To identify the owner of each account, the hash of agent $P$'s $pk_P$ is used to form an *address* [25]. The *address* is then used to receive transactions from other agents $P$. $Sk_A$ is used to sign a transaction with a valid signature from sending agent $A$ and sends it to another agent $B$'s address to verify it using sending agent $pk_A$. Hence, agent $P$ in the system can claim ownership of their account using $sk_P$ if their $sk_P$ is kept secret on their end devices and has not been shared with any malicious agent $P$.

The blockchain's built-in features, such as the digital signature, ensure that the proposed system's design goals DG 1 and DG 2 hold in the system. Below, we formalize all system authenticity properties.

- All sets of actions $\Gamma$ within agents $\lambda_P$. The authenticity of these actions is proven using Assumptions 1, 2, 5, and 7, which ensure that DG 1 and DG 2 hold in the system.
- All sets of actions $\Gamma$ beyond agents $\lambda_P$ are known as empty sequences, denoted by $\varepsilon$. The authenticity of $\varepsilon$ is proven using Assumptions 1, 3, 5, 6 and 7, which ensures that DG 3, DG 4 and DG 5 hold in the proposed system.

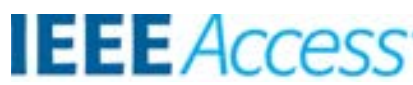

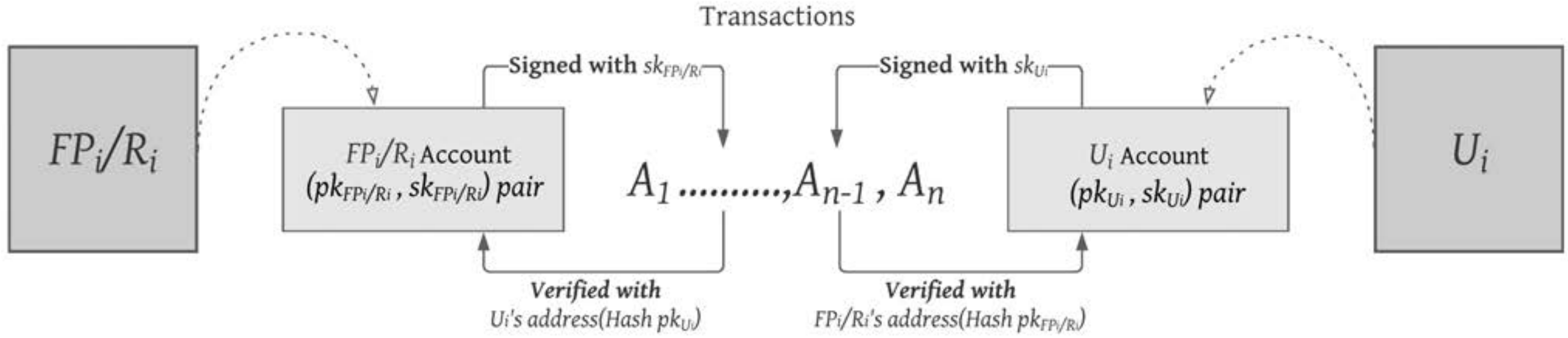

**FIGURE 10.** End-to-end authentication trust.

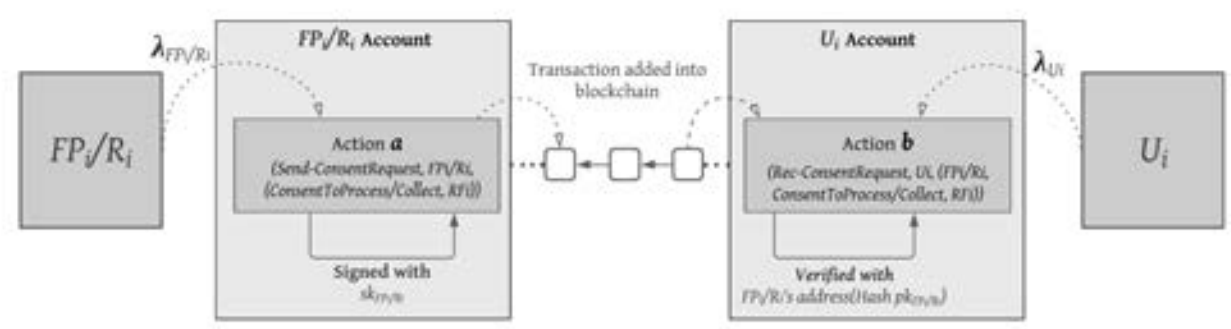

**FIGURE 11.** $\lambda_{U_i}$ for authenticity of send and received consent request actions.

#### 4) AUTHENTIC COMMUNICATION

Each agent in the system has a different $\lambda_P$ and can see only their own sent actions or received actions $A_1 \ldots A_n$. The authenticity of sent and received actions $A_1 \ldots A_n$ can be interpreted as *Auth(a, b, P)*. In other words, if a particular action ***b*** has occurred in a sequence of actions $\omega$, it must be authentic for agent $P$ that action ***a*** has occurred before. Therefore, to validate the authenticity of these actions, a precise definition of what ''authentic'' means is required. Hence, we use the SeMF's [18] definitions 1 and 2.

*Definition 1: A set of actions* $\Gamma \subseteq \Sigma$ *is authentic for* $P \in \mathbb{P}$ *after a sequence of actions* $\omega \in B$ *with respect to* $W_P$ *if* $alph(x) \cap \Gamma \neq \emptyset$ *for all* $x \in \lambda_P^{-1}(\lambda_P(\omega)) \cap W_P$.

Although Definition 1 allows sets of actions $\Gamma$ to be authentic, for the proposed system, a simplified definition for single actions is sufficient, as defined in the following definition from SeMF [18]:

*Definition 2: For a system S with behaviour* $B \subseteq \Sigma^*$, *agent* $P \subseteq \mathbb{P}$, *and actions* $a, b \in \Sigma$, $auth(a, b, P)$ *holds in B if for all* $\omega \in B$, *whenever* $b \in alph(\omega)$, *the action a is authentic for P.*

*Proposition 1: If agent B executes the ''received'' action (**b**), then the matching sent action (**a**) by agent A is authentic.*

**Note:** The reference number $RF_i$ in the following description is used to identify the consent required to process action $A_i$.

*Proof of Proposition 1:* In the proposed system, we can verify the authenticity of these actions using Assumptions 1, 2, 5, and 7. This proof shows the details for one ''receive'' action (**b**) in which the same proof also applies to all other receive actions in the proposed system. Fig. 11 shows that when agent A ($FP_i/R_i$) sends consent request transaction ***a***, its authenticity is verified by agent B ($U_i$) each time it executes the received action ***b*** (recvConsentRequest, $U_i$,($FP_i/R_i$, ConsentToProcess/Collect, $RF_i$)). Hence, the set of actions $\Gamma$ that will be authentic consists only of the send action ***a*** (sendConsentRequest, $FP_i/R_i$,(ConsentToProcess/Collect, $RF_i$)) and agent B ($U_i$), for which this action will be authentic. It is authentic for all sequences of actions $\omega$ that contain an action (recvConsentRequest, $U_i$,($FP_i/R_i$, ConsentToProcess/Collect, $RF_i$)).

As defined in Definition 1, it can be formally denoted as: For all sequence of actions $\omega \in S$ that (recvConsentRequest, $U_i$,($FP_i/R_i$, ConsentToProcess/Collect, $RF_i$)) $\in$ $alph(\omega)$ implies $alph(x) \cap$ {(sendConsentRequest, $FP_i/R_i$, (ConsentToProcess/Collect, $RF_i$))} $\neq \emptyset$ for all $x \in \lambda_{U_i}^{-1}(\lambda_{U_i}(\omega)) \cap W_{U_i}$.

This security property is provided by the proposed system with agent B's knowledge is equal to $W_B$ and agent B's *local view* equal to $\lambda_B$. $W_B$ is equal to Assumptions 1, 2, 5, and 7, and $\lambda_B$ is equal to received action ***b*** (recvConsentRequest, $U_i$, $FP_i/R_i$, ConsentToProcess/Collect, $RF_i$)).

To that end, we can validate Proposition 1 with agent B's knowledge $W_B$ equal to the blockchain's Assumptions 1, 2, 5, and 7:

- From Assumptions 1 and 7, we can conclude that if agent B receives action ***b*** $\implies$ , then this implies that the matching send action ***a*** sent by agent A is recorded in the blockchain.
- From Assumption 2, we can conclude that if send action ***a*** is added by agent A $\implies$ , then this implies that send action ***a*** is authentic.
- From Assumption 5, the action is authentic, meaning that the sent data are equal to the received data.

1 $\square$

Hence, the proposed system design goal DG 2 is met and can be captured using the following formalization:

$$auth(sendConsentRequest(ConsentToProcess/Collect, RF_i), recvConsentRequest(ConsentToProcess/Collect, RF_i), U_i) \quad (1)$$

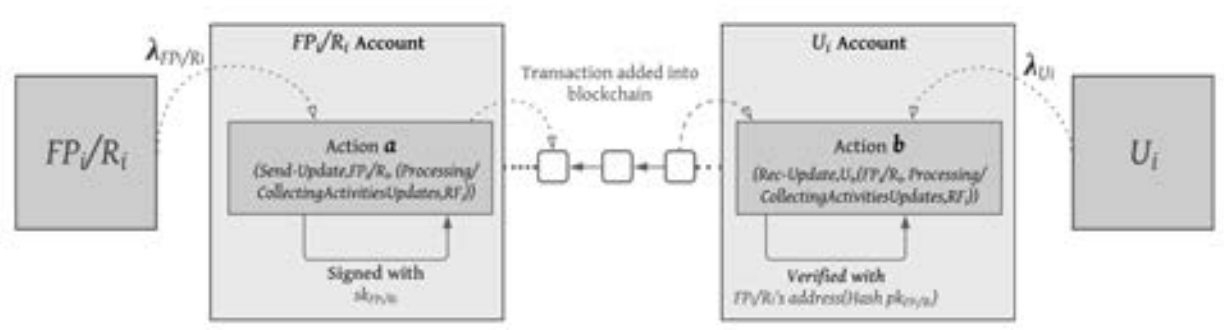

**FIGURE 12.** $\lambda_{U_i}$ **for authenticity of send and received update actions.**

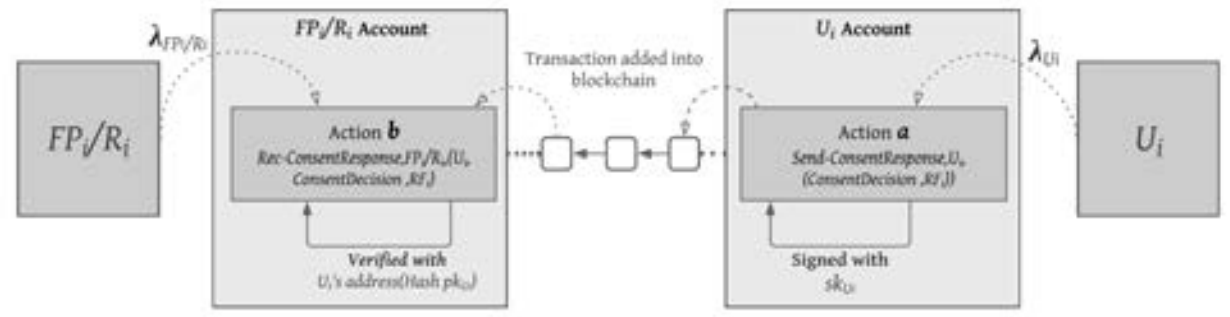

**FIGURE 13.** $\lambda_{FP_i/R_i}$ **for authenticity of sent and received consent response actions.**

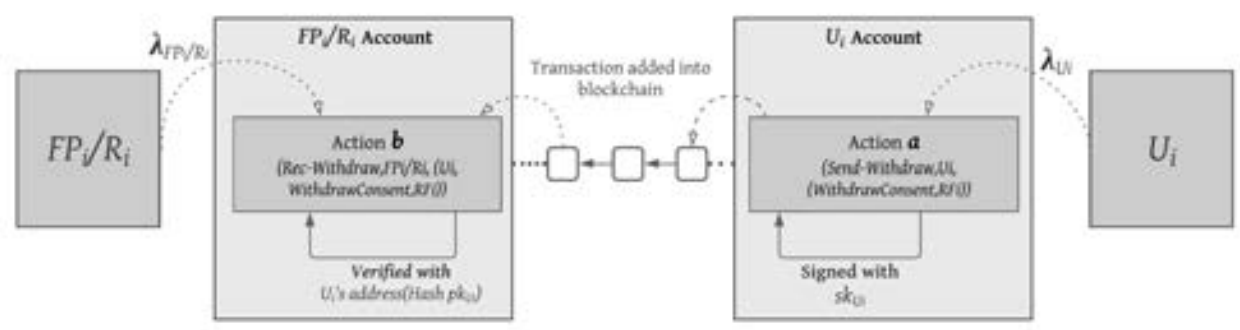

**FIGURE 14.** $\lambda_{FP_i/R_i}$ **for authenticity of sent and received withdrawal actions.**

We use the fact that Proof 1 applies to all other received actions, such as from $FP_i/R_i$ to $U_i$, from $U_i$ to $FP_i/R_i$, and from $FP_i$ to $R_i$, to satisfy DG 2. Below, we describe these actions as follows:

- Fig. 12 illustrates the authenticity of sending and receiving update actions from $FP_i/R_i$ to $U_i$, which can be formalized as follows.

$$auth(sendSelfReportDataAccess(Processing/CollectingUpdates, RF_i), recvSelfReportDataAccess(Processing/CollectingUpdates, RF_i), U_i) \quad (2)$$

- Fig. 13 shows the authenticity verification of the sent and received consent response actions, which can be formally stated as follows.

$$auth(sendConsentResponse(ConsentDecision, RF_i), recvConsentResponse(ConsentDecision, RF_i), FP_i/R_i) \quad (3)$$

- Fig. 14 depicts the authenticity verification of sent and received withdrawal actions and can be captured with the following formalization:

$$auth(sendWithdrawConsent(WithdrawConsent, RF_i), recvWithdrawConsent(WithdrawConsent, RF_i), FP_i/R_i) \quad (4)$$

However, there are some sets of actions $\Gamma$ in the proposed system beyond agents $\lambda_P$, known as empty sequences, which are denoted by $\varepsilon$. It is difficult to guarantee the authenticity of these actions $\varepsilon$ and prove that DG 3, DG 4, and DG 5 hold in the system. Below, we discuss the possible blockchain built-in mechanisms. Assumptions 6, 7, and 8 are used to ensure that DG 3, DG 4, and DG 5 hold in the system by addressing these difficulties. Blockchain's Assumptions 7 and 8 allow all sets of actions $\Gamma$ performed by agents $P$ to be recorded on the chain in a secure and reliable way and be viewable to all agents $\mathbb{P}$. This makes agent $P$ learn more about other actions and expands their $\lambda_P$ of the system actions beyond the send and receive actions.

A smart contract's reliability and inevitability features can fulfil the proposed system's design goals DG 3, DG 4, and DG 5. Assumption 6 effectively proves that these three design goals hold in the proposed system. As stated before in Definition 2, we state our propositions and prove them using Assumptions 1, 3, 5, 6, and 7.

*Proposition 2: If agent B executes the "received" action (**b**), then the matching sent action (**a**) by the smart contract is authentic.*

*Proof of Proposition 2:* Following Proof 1, we can also validate Proposition 2 with agent B's knowledge $W_B$ equal to Assumptions 1, 3, 5, 6, and 7.

- From Assumptions 1 and 7, we can conclude that if agent B receives action (***b***) $\Longrightarrow$ , then this implies that the matching send action (***a***) sent by the smart contract is recorded on the blockchain.
- From Assumptions 3 and 6, we can conclude that if send action (***a***) is added by the smart contract $\Longrightarrow$ , then this implies that send action (***a***) is authentic.
- From Assumption 5, an action being authentic means that the sent data equal the received data.

2 □

### 5) PROOF USING SMART CONTRACT $SC_1$

This process aims to ensure that the granted consent decision is valid for a period of time ***OR*** invalid if it has been withdrawn. Based on Proof 2, the proposed system design goal DG 3 holds in the system. This can be expressed as follows.

$$auth(auth(sendWithdrawConsen(WithdrawConsent, RF_i), recvWithdrawConsen(WithdrawConsent, RF_i), FP_i/R_i) \vee auth(t_v(expired), RF_i), \mathbb{P}) \quad (5)$$

### 6) PROOF USING SMART CONTRACT $SC_2$

This contract aims to check the validity of storing, processing, and collecting data based upon $U_i$'s granted/denied consent. Based on Proof 2, the proposed system design goal DG 5 holds in the system. This can be expressed as follows.

$$auth(auth(sendReportDataAvailable(ConsentToCollect, RF_i), recvReportDataAvailable(ConsentToCollect, RF_i), U_i) \wedge auth(sendConsentResponse(ConsentDecision, RF_i), recvConsentResponse(ConsentDecision, RF_i), FP_i/R_i) \wedge auth(SC_1 results, RF_i), \mathbb{P}) \quad (6)$$

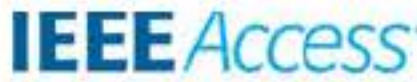

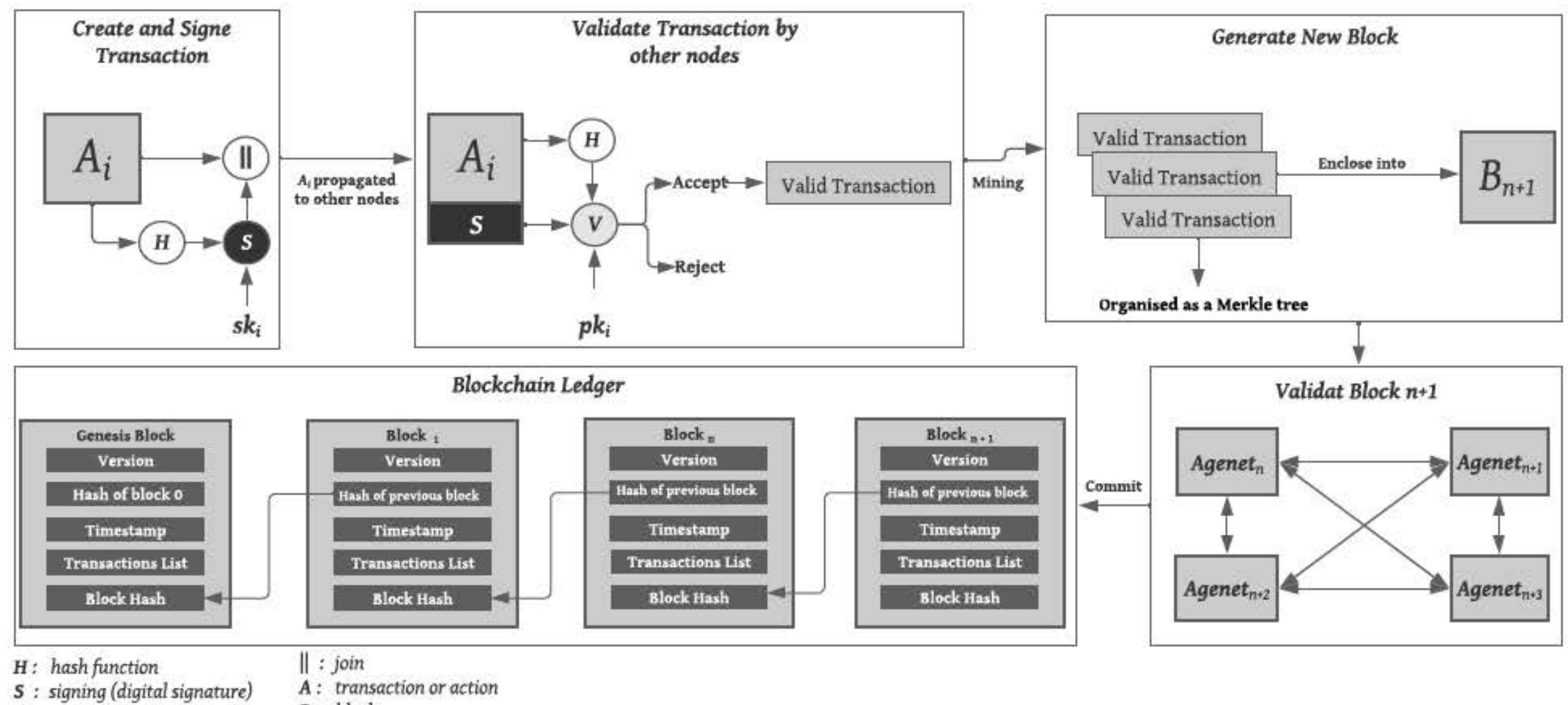

**FIGURE 15.** Blockchain's built-in approaches to guarantee nonrepudiation.

Similarly, DG 4 can be defined with $FP_i$'s reporting that $FP_i$ has accessed and processed fitness data from $U_i$'s fitness devices. This can be formalized as follows.

$$auth(auth(sendSelfReportDataAccess(ConsentToProcess, RF_i), recvSelfReportDataAccess(ConsentToProcess, RF_i), U_i) \wedge auth(sendConsentResponse(ConsentDecision, RF_i), recvConsentResponse(ConsentDecision, RF_i), FP_i/R_i) \wedge auth(SC_1 results, RF_i), \mathbb{P}) \quad (7)$$

Additionally, $SC_2$ can prove that the data collected by $R_i$ are valid and authentic by checking $R_i$'s report access to $FP_i$'s database. This process covers Proposition 2 and validates that DG 5 holds in the system. This can be formalized as follows.

$$auth(auth(sendSelfReportDataAccess(ConsentToCollect, RF_i), recvSelfReportDataAccess(c, RF_i), U_i) \wedge auth(sendConsentResponse(ConsentDecision, RF_i), recvConsentResponse(ConsentDecision, RF_i), FP_i/R_i) \wedge auth(SC_1 results, RF_i), \mathbb{P}) \quad (8)$$

Finally, we formalized all system actions using SeMF and proved that 1, DG 2, DG 3, DG 4, and DG 5, concerning authenticity and integrity, hold in the system.

### B. PROOF OF AUTHENTICITY

#### 1) DEFINITION OF THE PROPERTY

This property involves two pairs of communication in which both sent and received transactions cannot be denied. We use the definition from SeMF [18] to formalize the proof of the authenticity property introduced in Section IV:

*Definition 3 (Proof of authenticity): A pair $(\Gamma S, \Gamma P)$ with $\Gamma S \subseteq \Sigma$ and $\Gamma P \subseteq \Sigma$ is a pair of sets of proof actions of authenticity for a set $\Gamma \subseteq \Sigma$ on $S$ with respect to $(W_P)_{P \in \mathbb{P}}$ if for all $\omega \in S$ and for all $P \in \mathbb{P}$ with $alph(\pi_P(\omega)) \cap \Gamma P \neq \emptyset$ the following holds:*

*(1) For $P$ the set $\Gamma$ is authentic after $\omega$ and*

*(2) for each $R \in \mathbb{P}$ there exist actions $a \in \Sigma_{/P} \cap \Gamma S$ and $b \in \Sigma_{/R} \cap \Gamma P$ with $\omega ab \in S$.*

*Agent $P \in \mathbb{P}$ can give proof of authenticity of $\Gamma \subseteq \Sigma$ after a sequence of actions $\omega \in S$ if 1 and 2 hold.*

*Proposition 3: For the received actions (**b**), agent B must always be able to access evidence to show proof to other agents, which enables them to check the authenticity of the matching sent action (**a**) that occurred before the received action (**b**).*

#### 2) FORMALIZE ACTIONS PROOF OF AUTHENTICITY

Blockchain is an excellent solution to address this security property, which is used as an evidence recorder in which all transaction and activity logs of consent are submitted and settled permanently into a shared ledger. Based on Assumption 7, a blockchain is formed by a chain of blocks in which each block appends links to its direct predecessor until a block with index 0, which is known as the genesis block [25]. Based on Assumption 8, a copy of whole records is distributed among all entities, and it keeps chronologically nontamperable records that are linked via hash values. A digital signature scheme (asymmetric encryption), timestamp, and immutability of records, in which the system's actions can be tracked for authenticity proof, are all effective approaches to guarantee nonrepudiation [50]. Fig 15 depicts an overall view of blockchain's built-in approaches to guarantee nonrepudiation. For example, if there is any malicious modification in any block, all the blocks after that modified block must be recomputed because each block is tied to the hash value of the previous block's header, which proves that such malicious modifications cannot be performed in this case.

*Proof of Proposition 3:* As defined in Definition 3, we can validate Proposition 3 with agent B's knowledge $W_B$ equal to Assumptions 7 and 8.

- From Assumption 8, we can conclude that if agent B receives action ($\boldsymbol{b}$) $\implies$ , then agent B can access the evidence record.
- From Assumption 7, we can conclude that if agent B can access the evidence record $\implies$ , then agent B can provide proof that the sent action ($\boldsymbol{a}$) has occurred before the receive action ($\boldsymbol{b}$).

3 □

To that end, after formalizing the proposed system actions and in relation to the blockchain's built-in features, we guaranteed nonrepudiation by proving Proposition 3 using Proof 3.

## IX. CONCLUSION, LIMITATIONS AND FUTURE WORK

Sharing personal fitness data with third parties poses several privacy concerns. This study aims to address such concerns in the privacy policies of fitness trackers by improving user control over the processing of their fitness data. We designed a blockchain-based dynamic consent mechanism to mitigate privacy-preservation issues. The primary artifacts in the proposed system are the system architecture, requirements specification, and formal proof model. This study also evaluates the demonstrated artifacts using the SeMF tool and the identified blockchain assumptions. As a result, the SeMF evaluation proves that the system satisfies all our design goals. We conclude that the proposed system is suitable for preserving user privacy by improving user control over the processing of fitness data. However, further research is required to address the remaining gap: fitness trackers must honestly report their actual processing actions to the blockchain.

### A. LIMITATIONS

Although the proposed model has proven its effectiveness in terms of security properties using SeMF [18], there are some limitations to consider.

- The proposed framework requires an experimental evaluation to show whether the solution is practical to use in terms of the computation, storage, and communication overhead of each entity.
- This research provides only an abstract level of the proposed solution discussing the feasibility of utilising the blockchain to improve users' control over their data in fitness apps by consent mechanisms.
- The $FP_i$'s reported action in the proposed system still relies on $FP_i$ behaving and reporting honestly to the blockchain about the actual processing actions of $U_i$'s fitness data. Although the proposed system is based on the idea of decoupling consent management from data sharing, there must be a way to prove that the exchange of data occurred upon user decision making without relying on $FP_i$ being honest, as they own the database.

### B. FUTURE WORK

This study interprets the proposed system security requirements in verbal descriptions and then formalizes these requirements using the SeMF tool. The findings in this paper are just the first step in designing and developing blockchain-based consent management. In future work, we plan to conduct an experimental evaluation to demonstrate the effectiveness of the proposed model in terms of the computation, storage, and communication overhead of each entity. Additionally, further research is needed to determine the suitable choices of blockchain's technical details to the proposed solution, including but not limited to consensus algorithm, usage of tokens and their economy and the method for authenticating fitness devices.

## ACKNOWLEDGMENT

The authors acknowledge the efforts and contributions of Dr. Tharuka Rupasinghe (Ph.D., Monash University, and a Research Fellow at RMIT University), who gave tremendous support to this research.

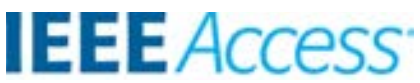

**MAY ALHAJRI** (Graduate Student Member, IEEE) received the B.Sc. degree (Hons.) in networking and telecommunication systems from the College of Computer and Information Sciences, Princess Nourah University, Riyadh, Saudi Arabia, in 2017, and the M.Sc. degree in cybersecurity from the Faculty of IT, Monash University, Melbourne, Australia, in 2021. She is currently a Teaching Associate with the Department of Computer Networks and Communications, College of Computer Sciences and Information Technology, King Faisal University, Al-Ahsa, Saudi Arabia. Before joining King Faisal University, she worked as an Industrial IoT Systems Engineer, in 2018.

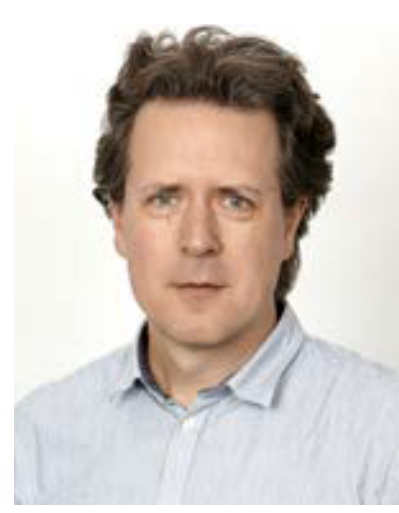

**CARSTEN RUDOLPH** received the Ph.D. degree from QUT, in 2001. He is currently an Associate Professor with the Faculty of IT, Monash University, where he heads the Department of Software Systems and Cybersecurity (SSC). He is the Deputy Director of the Blockchain Technology Centre, Monash University, and the Director of OCSC, Melbourne, Australia. Before joining Monash University, he was the Head of the Research Group on Trust and Compliance, Fraunhofer Institute for Secure IT, Darmstadt, Germany, and supported Huawei in setting up the Trusted Computing Research Laboratory, Germany. His research interests include information security, formal methods, security engineering, and cryptographic protocols.

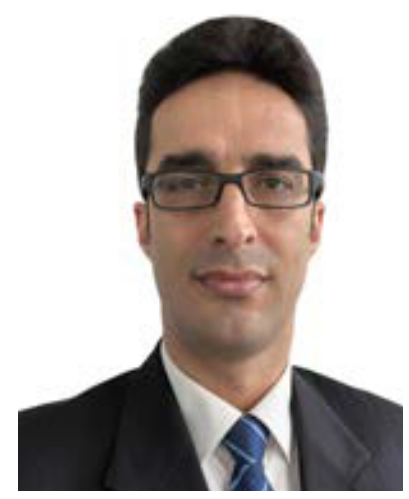

**AHMAD SALEHI SHAHRAKI** (Member, IEEE) received the M.Sc. degree in information security from the Faculty of Computing, UTM, in 2013, the M.Phil. degree in information security from the Science and Engineering Faculty, QUT, in 2017, and the Ph.D. degree in cybersecurity from the Faculty of IT, Monash University, and the DSS Group at CSIRO's Data61, in 2020. He held a RA position at the Cybersecurity Laboratory, Monash University, in 2021. He is currently a Research Fellow at the RMIT Blockchain Innovation Hub (BIH) and the Centre for Cyber Security Research and Innovation (CCSRI). His research interests include access control, blockchain, cryptography, cybersecurity, and healthcare.